\documentclass[preprint]{aastex}

\slugcomment{Accepted by AJ}

\newcommand{\eg}{\mbox{e.g.}}
\newcommand{\ie}{\mbox{i.e.}}

\newcommand{\skipthis}[1]{}

\newcommand{\be}{\begin{equation}}
\newcommand{\ee}{\end{equation}}
\newcommand{\e}{et al.\ }

\def\ie{{i.e.\ }}

\usepackage{epsf}
\usepackage{textcomp}

\begin{document}
\shorttitle{Near-IR Field Variable Stars in Cygnus OB7}

\title{Near-IR Periodic and Other Variable Field Stars in the Field of the Cygnus OB7 Star Forming Region}

\author{Scott J. Wolk}
\affil{Harvard--Smithsonian Center for Astrophysics, 60 Garden
Street, Cambridge, MA 02138}

\author{Thomas S. Rice}
\affil{Harvard--Smithsonian Center for Astrophysics, 60 Garden
Street, Cambridge, MA 02138}

\author{Colin A. Aspin}
\affil{Institute for Astronomy, University of Hawaii at Manoa, 640 N Aohoku Pl, Hilo, HI
96720}

\begin{abstract}
We present a subset of the results of a three season, 124 night, near-infrared monitoring campaign of the dark clouds Lynds 1003 and Lynds 1004 in the Cygnus OB7 star forming region.  In this paper, we focus on the field star population. Using three seasons of UKIRT J, H and  K band observations spanning 1.5 years, we obtained high-quality photometry on 9,200 stars down to J=17 mag, with photometric uncertainty better than 0.04 mag.  After excluding known disk bearing stars we identify 149 variables -- 1.6\% of the sample.  Of these, about 60 are strictly periodic, with periods predominantly $< 2$ days. We conclude this group is dominated by eclipsing binaries. A few stars have long period signals of between 20 and 60 days. About 25 stars have weak modulated signals, but it was not clear if these were periodic. Some of the stars in this group may be diskless young stellar objects with relatively large variability due to cool star spots. The remaining $\sim$60  stars showed variations which appear to be purely stochastic.

\end{abstract}


\section{Introduction}
\label{sec:Intro}

Two crucial experiments for identifying Young Stellar Objects (YSOs)  are near-infrared (NIR) disk studies and variability studies. We have combined these two techniques in a three season, 124 night NIR monitoring campaign of the star forming region Cyg OB7 (Rice \e 2012; hereafter Paper~I).   The first result of this experiment was the identification of 30 YSOs in the region and the discovery that about a quarter of them were seen to transition in JHK color-color space from the ``photospheric'' region of the NIR color-color diagram to the region of the diagram corresponding to disk bearing stars \citep{Lad92}.  In a follow-up paper, we study the periodic nature of the YSOs and the observed systematic color changes (Wolk \e 2013; hereafter Paper~III). 

Disk-bearing young stars, also known as classical T Tauri stars (cTTs), have long been identified as optically variable 
\citep{Joy45,Her62}.  At a minimum the variability is due to a combination of starspots, accretion and circumstellar disk occultations  \citep{Her94}.
The study of near-infrared variability in young stars allows us to directly study changes in those disk structures. Studies of the Orion A  and Chameleon I molecular clouds
established that variability is related to the presence of an inner accretion disk \citep{Car01, Car02}.
In Orion, as many as 93\% of the variable stars are identified to be young stars, and a strong connection is established between variability and near-infrared excess. 
Studies of individual YSOs such as AA Tau and its analogs reveal insights into magnetospheric accretion processes linked to inner-disk dynamics 
\citep{2003A&A...409..169B, 2007A&A...463.1017B, 2010MNRAS.409.1347D}.
Other types of stars such as EX Lup \citep{2010ApJ...719L..50A} and V1118 Ori \citep{2010A&A...511A..63A} exhibit large, eruptive mass-accretion events due to mass infall events of $M > 0.1 M_\oplus$  that are easily studied in the near-infrared; during outburst, their near-infrared emission is dominated by hotspot radiation and emission reprocessed in the disk.
Recent mid-IR variability surveys such in the IC 1396A and Orion star forming regions  
 provide key insights into physical processes of young stars over short ($\sim$40 day) timescales, finding that 70\% of YSOs with infrared excess are variable  \citep{Mor09, Mor11}.  \citet{Sch12} studied the NIR variability of several young clusters including the ONC, NGC 1333, IC 348 and $\sigma$ Orionis. He finds variability amplitudes are largest in NGC 1333, presumably because it has the youngest sample of YSOs. The frequency of highly variable objects also increases with the time window of the observations.

Although the focus of our observing campaign was to study YSOs, our observed field contains about 9200 background and foreground stars for which variability analysis were carried out.  Of these, 158 were found to be variable via the Stetson index \citep{Ste96}.  
Variable field stars provide important information about the nature and evolution of stars in various regions of our galaxy. For example, eclipsing binary stars provide us information about the masses and radii of stars \citep[see \eg][]{Hua56, Pop85}.  Pulsating variables, such as Cepheids and RR Lyrae, serve as distance indicators  \citep[see \eg][]{Ben07}.  
A common type of pulsators known as $\delta$ Scuti stars offer unique insight into the internal structure and evolution of main-sequence objects \citep{Tho03}.  \citet{Pie09} recently performed a deep four night monitoring campaign of about 50,000 stars in Carina with the VLT and found a 0.7\% variability fraction. The main goal of their program was the optical follow-up of OGLE planetary transit candidates. They found 43\% of the variables were eclipsing systems while a statistically identical number were pulsating systems.  About 2/3 of the pulsating systems were $\delta$ Scuti stars.

The goal of this paper is simple: to document the rate, and to a lesser extent the types, of NIR variability seen in the field stars. 
We will then use the result of the field star analysis as a point of comparison for the results found in star forming regions.
In the next section (\S 2) we will briefly review the source data. In \S 3 we discuss two techniques for period analysis: a fast $\chi^2$ minimization algorithm and Lomb-Scargle periodogram analysis. 
In 
\S 4 we present the results of our periodicity study of the field stars. The field stars divide into four types.  First -- regular periodic variables which are either eclipsing or pulsating;
second -- less regular periodic stars; third -- long duration variables which, despite a strong signal, defied description by a single period and fourth -- weak signal, stochastic variables. We summarize the results in \textsection \ref{sec:conclusion}.

\label{sec:data} 
\section{Observation and Data Reduction}

The constellation Cygnus contains many rich and complex star-forming regions, including the North America and Pelican nebulae and the Cygnus X star forming region \citep{2008hsf1.book...36R}. In the Cygnus region, nine OB associations have been found. Cygnus OB7 is the nearest at a distance of around 800 pc \citep[distance modulus $\mu$ = 9.5]{2009AJ....137..431A}. Cygnus OB7  is along the line-of-sight of  the large dark cloud Kh~141 \citep{Kha55}  -- also called the northern coal sack; this cloud is over 5$^{\rm {o}}$ across.  While a physical connection has not been confirmed, the dark clouds Lynds 1003-1004, the target of this study, lie near the middle of Kh~141. 
The dark clouds were first studied in the context of star formation by
\citet{Coh80}
who found a diffuse red nebula he named RNO 127.
This nebula was later determined to be a bright Herbig-Haro (HH) object by 
\citet[HH 428][]{Mel01,Mel03}.
Further study in the optical and near-infrared identified a number of Herbig-Haro objects 
\citep{Dev97,Mov03}
and multiple IRAS sources 
\citep{Dob96}
that reveal the presence of a young stellar population and significant star formation activity.
 The field of view of the study is about a 1 degree square centered near 21h00m +52$^{\rm o}$30\arcmin\ (J2000.) near the ``Braid Nebula Star''.  While \citet{2009AJ....137..431A} and \citet{Ric12} have identified over 40 disked YSOs in the region, to date no deep X-ray studies have been made to identify the disk-free young stellar population.

The dataset used here was fully described in Paper~I.  In brief,  $J$, $H$, and $K$ observations of the Cygnus OB7 region were obtained using the Wide Field Camera (WFCAM) instrument on the United Kingdom InfraRed Telescope (UKIRT), an infrared-optimized 3.8 meter telescope atop Mauna Kea, Hawaii at 4200 meters elevation. Our data consist of WFCAM observations taken from May 2008 to October 2009 in three observing seasons as part of a special observation program.

Data were taken on a total of 124 nights during this period.  
For each night, we extracted data from the archive for all stars with photometric uncertainties less than 0.1 mag in J  band and no processing error flags. 
Starting from about 100,000 detected objects, over 38,000 stars met this criterion every night.
The typical errors of these stars for one night are shown in  Figure~1.
Out of the 124 nights in the original survey, 24 nights were rejected due to deviations in the mean color of $\gtrsim 1\%$, leaving 100 nights for our analysis (Table~1) 
\citep[see also][]{Ric12}.
To assure the best quality data we  require $J \lesssim 17.3 $ which limits errors to $\le 4\%$ at J; we also excluded stars brighter than  J =11 as bright stars would saturate in epochs with especially good seeing or if the star brightened.  We also required no quality error bit flags on any of the observations.
In the end we obtained high-quality photometry on 9,200 stars.

\begin{figure}[htbp]
\begin{center}
\includegraphics[scale=1.2]{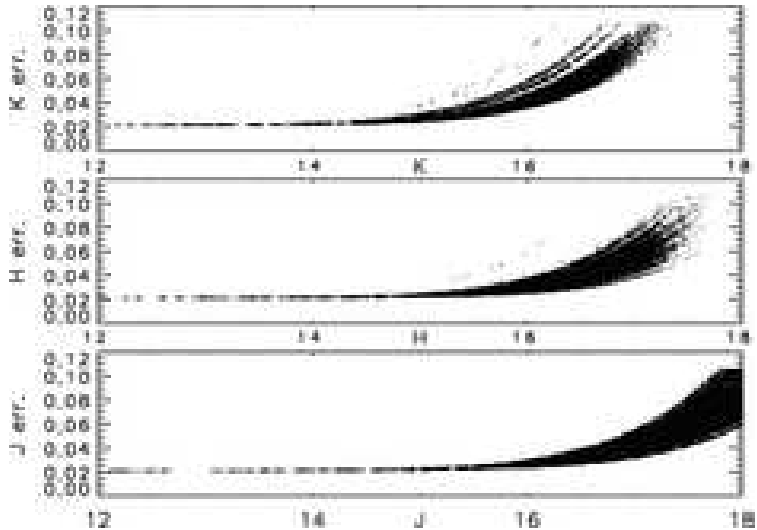}
\caption{Errors vs. magnitude in all three bands shown for one night of data. This is about 38,000 stars with $J$ errors $<$ 0.1 mag.}
\end{center}
\label{fig:err}
\end{figure}

\begin{deluxetable}{lrrr}
\tablewidth{0pt}
\tablecaption{Observing Log} 
\tablehead{
\colhead{Epoch}
  &\colhead{Start Date}
    &\colhead{End Date}
      &\colhead{Number of observations}
      }
\startdata
Season 1&  26-Apr-2008  &  22-May-2008 &  19    \\
Season 2 & 19-Sep-2008 &  27-Nov-2008  &  39   \\ 
Season 3 & 28-Aug-2009 &  13-Oct-2009   &  41   \\
\hline
Total         & 26-Apr-2008 &    13-Oct-2009  &  100\\
\enddata
\label{log}
\end{deluxetable}

Variability was quantified for all sources in the field using the Stetson index ($S$; Stetson 1996)\footnote{This index is referred to by the letter $J$ in the original article, we use $S$ to avoid confusion with the filter band.}.  This is a method for quantifying variability within a sample which includes multiple colors each with different error characteristics.  The resultant value is zero for a constant source and exceeds 1 for a source with significant correlated variability; high values indicate greater variability.\footnote{\citet{Car01} compared the Stetson index to $\chi^2$ fits and concluded that $S>0.55$ was sufficient to confirm variability.  In Paper~I we concluded that $S>1.0$ was preferable for our analysis because the more conservative value prevented false detections in large samples.}   

\label{sec:analysis}
\section{Period Analysis}

\subsection{Automated Period Searches}
To investigate whether the variable stars were periodic, we used two period-finding algorithms: the Lomb-Scargle periodogram \citep[also knonwn as the Lomb Normalized Periodogram -- LNP; cf.][]{1989ApJ...338..277P} and the Fast $\chi^2$  (F$\chi^2$) algorithm \citep{2009ApJ...695..496P}, both suited for period analysis on unevenly sampled data such as ours. The Lomb-Scargle method is a useful and popular way to analyze periodicity with an easily interpreted periodogram, which identifies multiple candidate periods and their relative probability. This method essentially takes the Fast Fourier Transform of the input signal and reinterprets it (with certain restrictions) as the  periodogram. The signal in the periodogram with the maximum power is generally interpreted as the true signal. The method uses the number and spacings of samples to determine the frequency domain to be explored.  Spurious detections can be produced at frequencies where the sample times have significant periodicity (\ie nightly samples); power from harmonics above the fundamental sine wave is lost, and the algorithm ignores point-to-point variations in the measurement error.

\begin{deluxetable}{lrrr}
\tablewidth{0pt}
\tablecaption{Period search parameters.} 
\tablehead{
\colhead{Description}
  &\colhead{Parameter}
      }
\startdata
J Range    &  $\sim$11.0-17.3 mag\\
H Range    &   $\sim$11.0-16.5 mag\\
K Range    &   $\sim$11.0-16.0 mag\\
Typical errors & 0.02 mag \\
Stars Monitored  & $\sim$ 9,200\\
Range of sensitivity & 0.1- 250 days\\
Range of completeness  & 0.2 - 50 days \\ 
Amplitude Limit & 0.02  \\
Typical $\sigma_{Period} $&  $< 0.01$ days \\
\enddata
\label{tab:aana}
\end{deluxetable}

Given the long baselines in our data, the LNP method can determine very precise periods. The formal error on
derived (linear) frequencies is: $\delta f =  3 \sigma_N / 2TA\sqrt{N_0}$ \citep{Hor86},
where  $\sigma_N$
 is the variance of noise remaining after the periodic signal is subtracted, $N_0$ is the number of independent points (about 100), $T$ is the length of time baseline (about 500 days), and $A$ is the amplitude of signal (typically 10\%).  For our data, the formal error is typically less than 0.001 days,  an indication of the tremendous leverage gained by long term  monitoring of very stable periods. On the other hand the step size of the period search algorithms is a function of the square of the period and while the step size is $\sim$ 0.001 day for periods $< 2$ days, the step size is about 0.014 days for a period of 10 days.  We list measures of period to the nearest 0.01 days ($\sim$ 15 minutes) which is well within the precision for periods $<$ 10 days and reasonable for all listed periods. This precision is consistent with our purposes,  which is to identify the field variables, not to physically evaluate each one in detail.

Among non-variables in our data, our LNP analysis often found spurious periods near 0.5 and 1.0 days, likely due to the $\sim$ 1 day separation between nightly observations.  
Because of our sampling rate and the induced aliasing in our dataset, we chose to reject certain windows of periods from our LNP analysis. These were periods shorter than 0.55 days, longer than 250 days, and the interval 0.9 - 1.1 days.  Further, the LNP method favors sinusoidal variability and had a difficulty with sharp eclipsing systems. This happens because the Fourier transform is at the core of the signal decomposition. Hence the signal is decomposed into sines and cosines, no matter what the shape of the true signal.   

To overcome these limitations, we also used the  F$\chi^2$ algorithm. 
The F$\chi^2$ method uses a truncated Fourier expansion.  Since a limited number of harmonics are explored, it allows the frequencies searched to have arbitrary range and density so that periods are not missed by the search.  F$\chi^2$ is sensitive to power in higher harmonics of the fundamental sinusoid  meaning it can identify structures more complex than sinusoids. 
Further, there is  insensitivity to the sample timing including periods less than one day and near one day. 
The frequency search is gridded more tightly than the traditional integer number of cycles over the span of observations, eliminating power loss from peaks that fall between the grid points.  Using the  (F$\chi^2$) algorithm  we were able to  explore periods as short as 0.1 days and found very stable results across the three bands.  However, simulations with test data indicated that period near 0.1 days still tended to get lost in aliasing with the sampling rate, especially in the single season data.
Empirically, the F$\chi^2$ method is more reliable for complex variables such pulsating stars and for short-period ($\lesssim$ 1 day), highly stable, but non-sinusoidal variability, such as detached eclipsing binaries.  Table~2 
lists the overall parameters of the periodicity search.

Data from each filter and each season were independently analyzed by both the F$\chi^2$ and LNP techniques.  This analysis was followed by analysis of the combined three epochs of data (4 seasonal divisions $\times$ 3 filters $\times$ 2 methods = 24 light curves generated for each star).  Results of the combined data set were given primacy as dynamic periods, either due to eclipses or pulsations, should be stable over 500 days. The individual seasonal measurements were more important for the pre-main sequence stars in which the mechanisms inducing the periodicity, whether cool star spots, accretion spots or disk features, are thought to be only quasi-stable in nature. 

\subsection{Manual Validation}
Even when it appears a period is present, the algorithms often disagree on the preferred period.   We manually graded the light curves of each star on a 5 point empirical scale. A grade of 5 was given to the subset of periods which appeared beyond question; usually these were clear eclipsing systems or other systems which varied sinusiodally with periods under a day.
A grade of 4 was given to stars with periods of similar consistency in results, but in which the noise in the individual measurements was  nearly at the level of the signal and hence less reliable.  A grade of 3  was given when the various techniques returned clear aliases of each other an hence the positive determination of the true period was not possible.   Periods longer than 60 days often fall into this category.  A grade of 2 was given when the various methods returned different periods of similar quality and hence the period list was not considered to be reliable.  A grade of 1 was assigned to cases in which the rise and fall in the data were clear, but no period could be established, perhaps simply due to unlucky sampling.    Finally, a 0 is used for stars which are variable, but have no evidence of periodicity. 

\begin{figure}
\includegraphics[scale=1.0, trim = 0mm  34mm 0mm 0mm ,clip]{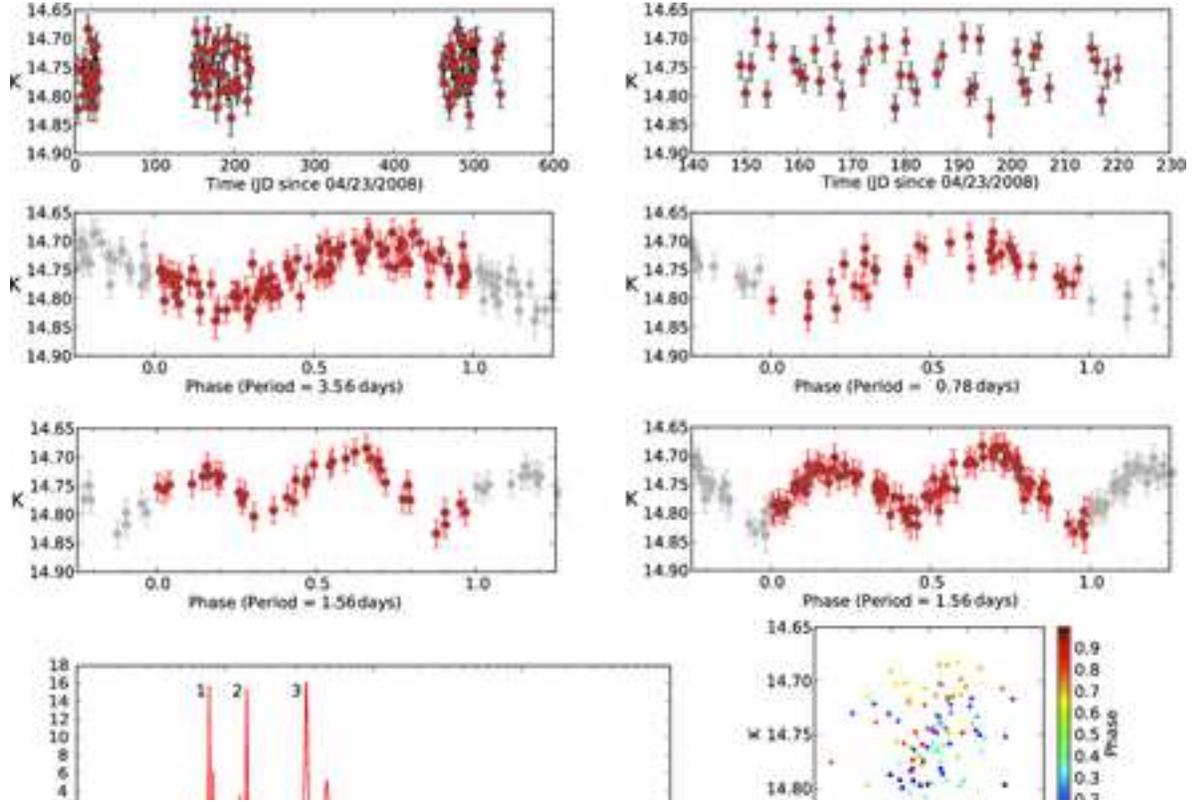}

   \caption{An example period analysis for  205928.56+521010.4.  
   Top-left: The raw  $K$-band lightcurve for all three seasons.  Top-right: The raw  $K$-band lightcurve for season 2 shown for clarity.
   Second-left:  The result of folding over the strongest signal in the $K$-band LNP using data for all three seasons LNP is best suited to finding single sinusoidal periods.
   Second-Right: The result of folding over the strongest signal in the $K$-band data found in the  $\chi^2$ minimization data,  this is also one of the harmonics in the LNP (bottom-left). Note the RMS on the curve exceeds the nominal error.
    Third-Left: Now folded on the intermediate harmonic -- season 3 only.  Double sine wave pattern is evident with very low residuals. 
    Third-Right: The result of folding all $K$-band data over 1.56 days. 
    Note that the peaks are unequal indicative of an eclipsing contact binary system. 
       Bottom-left: The periodogram of the $K$-band data for all three seasons, the peaks labeled ``1'',``2'' and ``3'' refer to the peaks of 
       0.78, 1.56, 3.56 days respectively.
        Bottom-right: Phased color-magnitude diagram shows little color dependence as a function of brightness or phase.}
     \label{Fig:V102}
\end{figure}

 For the intermediate cases, Figure~\ref{Fig:V102} shows an example of the follow-up analysis. The individual periodograms show multiple peaks. In this case, they are near 0.78 days, 1.55 days and 3.55 days.  The peaks are sharpest in the combined three season data, indicating a very stable long term signal. Though the signal peaks are similar in size, inspection of the three fits shows the 1.55 day period has the smallest deviations from a smooth fit to the data.  Further, we note, in the 1.55 day decomposition there are two peaks of different sizes and hence when they are folded on top of each other the apparent noise is enhanced.  This is a contact eclipsing system, similar to W~UMa.

This approach does not work when the data are too symmetric.
Many of the minimum $\chi^2$ phase-folded light curves showed a single fall and rise (see the third panel of Figure~2).  This is surprising as we expected eclipses to be common but most eclipse lightcurves have both a primary a secondary eclipse.  In addition, when we compared the periods of systems with a single rise fall to those that showed two, the periods of the single cycle objects where shorter than those which showed both a primary a secondary eclipse  -- this indicated a bias in the period finding.   Formally,  the difference in $\chi^2$ between a lightcurve showing a single event and dual events very small.  Hence the period finding codes will tend to favor, incorrectly, the shorter period showing a single eclipse. 
Because of this, we manually reviewed all periodic sources and doubled the period of those which indicated a single eclipse.  
We corrected only systems that  showed highly stable periods consistent with eclipse induced variations.  We do not apply the correction to more stochastic systems since other phenomena, especially pulsation and star spots, could induce a periodic single dip modulation.

\section{Results}
\label{sec:results}

We divided the field variable stars into four groups based on their periodic signatures.   
About 60 
appeared to be eclipsing systems with strong regular variations on time scales usually less than two days. 
Around 20 had weak modulation on the order of a few days -- consistent with star spots. Three appear to have long term cycles of between 20 and 60 days.  
The remainder, about 60 variables, showed no signs of periodic behavior. We labeled these stochastic variables. About 90\% of these had $S < 2$ and none had $S > 2.7$.
	A cursory examination found 17 lay along the Lynds~1003-1004 dust ridge, consistent with a random spatial distribution.  There was an over abundance of the stochastic sources south of the dust lane, compared to north,  but this was not found to be statistically significant.

In Tables~3--5,
we quantify the variability observed in these sources. The first two columns of these tables list the position of the sources.
Later, we will refer to particular star by the position, contracted by the removal of the colons. 
Columns 3-5 list the median of the 10 brightest measurements of the star in the respective filters, $J$, $H$ and $K$.  This gives the relevant brightness of the star out of eclipse while minimizing observational biases.   
Columns 6 and 7 list the range seen in the $J$ and $K$ filters. Similarly, 
columns 8 and 9 list the peak to floor change in the  $J-H$ and $H-K$ colors.  Column 10 lists the Stetson index for the star.  
Tables~3 and 5  
have an eleventh column listing the period of the star. Table~3 
adds a final column to indicate the {\sl classification} of the binary  which we discuss in \S 4.1.
Table~5 
adds a final column to indicate the {\sl grade} of that period which we discussed in \S 3.2.



\begin{deluxetable}{rllllllccrrr}

\tabletypesize{\scriptsize}
  
  \tablecaption{Strongly periodic stars in the monitored field.}
  \label{Table:VAR1}
  \tablewidth{0pt}

 \tablehead{
    \colhead{R.A. } & 
    \colhead{Decl.} & 
    \colhead{$J$ } &
    \colhead{$H$ } &
    \colhead{$K$ } &
    \colhead{$\Delta J$ } &
    \colhead{$\Delta K$ } &
    \colhead{$\Delta J-H$ } &
    \colhead{$\Delta H-K$ } &
    \colhead{$S$}  &
\colhead{Period} &
\colhead{Type} \\
   \colhead{(J2000)} &     \colhead{(J2000)} & 
    \colhead{mag} &
    \colhead{mag } &
    \colhead{mag } &
    \colhead{mag} &
    \colhead{mag } &
    \colhead{mag } &
    \colhead{mag} &
    \colhead{~}  &
\colhead{(days)}&
    \colhead{~} 
  }
\startdata
{\bf Simple} &  {\bf eclipse}  &  &  &  &  &  &  &  &  &  \\
\hline
  20:58:25.76 & +52:22:48.0 & 16.98 & 15.54 & 14.83 & 0.30 & 0.19 & 0.15 & 0.14 & 1.69 & 26.34 & EA\\ 
  20:58:52.56 & +52:26:23.5 & 15.33 & 14.08 & 13.33 & 0.21 & 0.18 & 0.08 & 0.07 & 3.84 & 1.80& E\\ 
  20:59:08.17 & +52:45:31.6 & 16.70 & 15.82 & 15.38 & 0.80 & 0.74 & 0.39 & 0.31 & 10.12 & 0.92& E\\ 
  20:59:14.13 & +52:53:30.6 & 15.90 & 15.20 & 14.83 & 0.18 & 0.17 & 0.09 & 0.10 & 2.19 & 2.36& E\\ 
  20:59:26.53 & +52:52:47.3 & 14.04 & 13.49 & 13.17 & 0.24 & 0.23 & 0.08 & 0.06 & 4.40 & 3.90& E\\ 
  20:59:39.90 & +52:03:43.4 & 14.68 & 14.08 & 13.85 & 0.16 & 0.16 & 0.06 & 0.07 & 2.09 & 1.10& E\\ 
  20:59:56.53 & +52:48:44.4 & 16.24 & 15.35 & 14.91 & 0.79 & 0.70 & 0.27 & 0.22 & 12.61 & 0.76& E\\ 
  21:00:33.27 & +52:14:24.6 & 16.58 & 15.52 & 14.95 & 0.55 & 0.56 & 0.16 & 0.17 & 6.92 & 2.72& E\\ 
  21:00:43.81 & +52:13:02.2 & 16.84 & 15.32 & 14.57 & 0.44 & 0.25 & 0.25 & 0.10 & 4.58 & 4.20& EB\\ 
  21:00:51.23 & +52:50:35.5 & 15.43 & 14.75 & 14.39 & 0.45 & 0.45 & 0.10 & 0.09 & 5.48 & 6.84& E\\ 
  21:01:49.79 & +52:52:35.9 & 16.02 & 15.24 & 14.91 & 0.92 & 0.78 & 0.20 & 0.21 & 18.89 & 0.74& EB\\ 
  21:02:18.28 & +52:26:08.4 & 14.81 & 14.18 & 13.90 & 0.29 & 0.30 & 0.10 & 0.10 & 2.13 & 0.76& E\\ 
  21:02:23.06 & +52:50:56.6 & 16.73 & 16.04 & 15.59 & 0.52 & 0.63 & 0.20 & 0.24 & 5.87 & 1.94& EA\\ 
  21:02:40.50 & +52:43:52.3 & 16.71 & 15.95 & 15.53 & 0.69 & 0.42 & 0.28 & 0.28 & 4.19 & 1.36& E\\ 
  21:03:10.79 & +52:22:04.3 & 15.68 & 14.61 & 13.96 & 0.45 & 0.47 & 0.08 & 0.08 & 9.16 & 2.40& E\\ 
  21:03:16.39 & +52:14:48.6 & 16.42 & 15.59 & 15.13 & 0.58 & 0.64 & 0.14 & 0.15 & 7.71 & 3.30& E\\ 
  21:03:19.74 & +52:40:26.2 & 13.26 & 12.77 & 12.57 & 0.15 & 0.24 & 0.12 & 0.15 & 2.36 & 1.56& E\\ 
{\bf Continual} & {\bf change}   &  &  &  &  &  &  &  &  & \\
\hline
  20:57:40.61 & +52:30:38.5 & 15.87 & 15.07 & 14.72 & 0.42 & 0.40 & 0.17 & 0.15 & 10.56 & 0.26& EW\\ 
  20:57:42.55 & +52:49:22.3 & 15.73& 14.92 & 14.43 & 0.20 & 0.23 & 0.21 & 0.19 & 5.69 & 1.11& EW\\ 
  20:57:56.91 & +52:40:19.7 & 15.73 & 15.03 & 14.69 & 0.32 & 0.33 & 0.10 & 0.11 & 9.78 & 0.38& EW\\ 
  20:58:01.46 & +52:22:37.2 & 16.09 & 15.45 & 15.14 & 0.21 & 0.23 & 0.19 & 0.13 & 4.28 & 0.28& EW\\ 
  20:58:01.70 & +52:20:09.3 & 13.54 & 13.14 & 12.93 & 0.14 & 0.13 & 0.08 & 0.07 & 3.19 & 0.30& EW\\ 
  20:58:14.05 & +52:47:24.7 & 15.06 & 14.44 & 14.10 & 0.24 & 0.34 & 0.18 & 0.08 & 7.94 & 0.93& EW\\ 
  20:58:15.20 & +52:53:57.8 & 17.26 & 16.20 & 15.76 & 0.51 & 0.43 & 0.40 & 0.30 & 5.19 & 0.20& EW\\ 
  20:58:42.87 & +52:44:29.6 & 16.71 & 15.72 & 15.31 & 0.90 & 0.79 & 0.35 & 0.33 & 14.37 & 0.28 & EB\\ 
  20:59:05.47 & +52:07:11.3 & 15.38 & 14.72 & 14.39 & 0.32 & 0.31 & 0.10 & 0.07 & 9.58 & 0.50& EW\\ 
  20:59:38.03 & +52:17:02.5 & 16.88 & 15.42 & 14.61 & 0.46 & 0.35 & 0.19 & 0.13 & 9.72 & 0.44& EW\\ 
  20:59:42.46 & +52:37:40.0 & 17.00& 16.14 & 15.72 & 0.56 & 0.45 & 0.25 & 0.17 & 8.15 & 0.38& EW\\ 
  20:59:45.69 & +52:07:16.3 & 15.96 & 15.26 & 14.98 & 0.42 & 0.42 & 0.14 & 0.19 & 13.19 & 0.36& EW\\ 
  20:59:51.01 & +52:03:48.1 & 14.98 & 14.19 & 13.76 & 0.15 & 0.15 & 0.07 & 0.05 & 3.32 & 0.92& EW\\ 
  20:59:53.48 & +52:08:09.3 & 15.09& 14.52 & 14.34 & 0.12 & 0.15 & 0.09 & 0.10 & 2.92 & 0.30& EW\\ 
  21:00:29.43 & +52:11:47.1 & 17.02 & 15.31 & 14.36 & 0.42 & 0.26 & 0.23 & 0.09 & 6.77 & 0.40& EW\\ 
  21:00:37.56 & +52:17:53.9 & 15.69 & 14.77 & 14.36 & 0.23 & 0.23 & 0.10 & 0.09 & 7.30 & 0.28& EW\\ 
  21:00:38.51 & +52:31:46.1 & 17.10 & 15.94 & 15.31 & 0.35 & 0.30 & 0.33 & 0.23 & 5.57 & 0.36& EW\\ 
  21:01:07.03 & +52:24:41.8 & 14.12 & 13.12 & 12.63 & 0.32 & 0.26 & 0.08 & 0.08 & 9.71 & 0.43& EW\\ 
  21:01:27.73 & +52:42:55.8 & 16.70 & 15.72 & 15.27 & 0.40 & 0.33 & 0.20 & 0.23 & 7.31 & 0.40& EW\\ 
  21:01:31.13 & +52:47:10.0 & 16.22 & 15.36 & 14.90 & 0.24 & 0.17 & 0.17 & 0.09 & 2.89 & 0.42& EW\\ 
  21:01:35.18 & +52:09:16.7 & 16.40 & 15.24 & 14.66 & 0.40 & 0.35 & 0.20 & 0.14 & 9.03 & 0.40& EW\\ 
  21:01:45.39 & +52:40:59.9 & 14.81 & 14.15 & 13.92 & 0.33 & 0.34 & 0.15 & 0.15 & 11.90 & 0.28 & EA \\  
  21:02:01.04 & +52:43:13.8 & 16.84 & 15.85 & 15.42 & 0.35 & 0.34 & 0.20 & 0.17 & 6.04 & 0.24& EW\\ 
  21:02:09.77 & +52:47:29.4 & 15.75 & 14.83 & 14.40 & 0.24 & 0.24 & 0.11 & 0.10 & 5.20 & 1.40& EW\\ 
  21:02:09.93 & +52:31:35.4 & 17.11 & 16.19 & 15.59 & 0.29 & 0.23 & 0.24 & 0.16 & 2.60 & 1.04& EB\\ 
  21:02:24.97 & +52:46:14.0 & 16.86 & 16.03 & 15.64 & 0.55 & 0.54 & 0.40 & 0.35 & 10.88 & 0.32& EW\\ 
  21:02:31.31 & +52:11:08.9 & 16.04 & 15.13 & 14.72 & 0.34 & 0.30 & 0.12 & 0.11 & 8.84 & 0.36& EW\\ 
  21:02:40.60 & +52:34:48.4 & 12.39 & 11.86 & 11.57 & 0.49 & 0.45 & 0.11 & 0.07 & 16.91 & 0.94& EW\\ 
  21:03:05.18 & +52:55:01.1 & 16.86 & 15.96 & 15.57 & 0.43 & 0.31 & 0.24 & 0.35 & 4.14 & 0.24& EW\\ 
{\bf Other} & {\bf periodic} &  &  &  &  &  &  &  &  &  &  \\
\hline
  20:58:24.90 & +52:30:31.5 & 17.19 & 15.15 & 14.12 & 0.74 & 0.42 & 0.37 & 0.13 & 8.10 & 10.12 & P \\ 
  20:59:31.39 & +52:28:02.1 & 16.37 & 15.33 & 14.76 & 0.48 & 0.33 & 0.26 & 0.15 & 9.02 & 0.61 & RRab \\  
  21:00:02.10 & +52:17:26.1 & 16.57 & 15.52 & 14.91 & 0.17 & 0.12 & 0.14 & 0.13 & 1.61 & 23.08 & EA?\\   
  21:02:05.76 & +52:22:53.9 & 16.21 & 14.04 & 12.92 & 0.24 & 0.16 & 0.13 & 0.08 & 4.70 & 24.29 & P? \\   
  21:02:23.22 & +52:49:51.0 & 16.34 & 15.55 & 15.08 & 0.64 & 0.54 & 0.17 & 0.21 & 8.75 & 0.25& $\delta$ Scu\\ 
{\bf Lower}  & {\bf quality}  &  &  &  &  &  &  &  &  & \\
\hline  
 20:57:46.71 & +52:03:02.3 & 14.13 & 13.53 & 13.28 & 0.09 & 0.11 & 0.06 & 0.08 & 1.85 & 0.40& EW\\ 
  20:57:49.45 & +52:36:13.6 & 16.64 & 15.81 & 15.33 & 0.21 & 0.22 & 0.18 & 0.20 & 1.30 & 3.52& E \\ 
  20:57:50.41 & +52:33:28.0 & 13.02 & 11.93 & 11.39 & 0.28 & 0.28 & 0.09 & 0.08 & 1.52 & 0.16& E\\ 
  20:58:20.99 & +52:23:00.8 & 16.90 & 16.09 & 15.66 & 0.29 & 0.23 & 0.23 & 0.22 & 1.42 & 1.72& E\\ 
  20:58:21.17 & +52:54:45.1 & 16.43 & 15.63 & 15.28 & 0.25 & 0.23 & 0.16 & 0.19 & 3.99 & 0.29& EW\\ 
  20:58:33.79 & +52:30:33.0 & 16.22 & 14.32 & 13.12 & 0.20 & 0.11 & 0.15 & 0.07 & 2.36 & 2.19& EW\\ 
  20:59:22.81 & +52:23:38.4 & 15.59 & 14.67 & 14.20 & 0.09 & 0.08 & 0.08 & 0.08 & 2.10 & 0.40& EW\\ 
  20:59:28.88 & +52:45:44.1 & 15.20 & 12.89 & 11.50 & 0.26 & 0.18 & 0.09 & 0.08 & 2.43 & 2.06& E\\ 
  20:59:36.91 & +52:54:11.1 & 16.08  & 15.13 & 14.73 & 0.15 & 0.09 & 0.14 & 0.09 & 1.18 & 2.93& E\\ 
  21:00:54.62 & +52:32:13.4 & 16.53 & 15.59 & 15.01 & 0.39 & 0.28 & 0.39 & 0.18 & 2.31 & 8.12& E\\ 
  21:00:58.93 & +52:38:10.7 & 14.72 & 13.97 & 13.56 & 0.13 & 0.11 & 0.12 & 0.12 & 1.61 & 0.38& EW\\ 
  21:01:22.54 & +52:37:52.0 & 16.30 & 15.55 & 15.10 & 0.40 & 0.33 & 0.28 & 0.13 & 4.56 & 1.64& EA\\ 
 21:01:58.81 & +52:50:41.6 & 15.62 & 14.94 & 14.59 & 0.10 & 0.12 & 0.08 & 0.08 & 2.27 & 0.36& EW\\ 
  21:02:44.91 & +52:42:51.3 & 16.20 & 15.31 & 14.92 & 0.21 & 0.20 & 0.14 & 0.15 & 4.46 & 0.16& EW\\ 
\enddata

  \tablecomments{Typical photometric errors are $\sim2\%$}

\end{deluxetable}



\begin{deluxetable}{lllllllccl}

\tabletypesize{\scriptsize}
  
  \tablecaption{Stochastically variable stars in the monitored field.}\label{Table:stoc}
  \tablewidth{0pt}

 \tablehead{
     \colhead{R.A. } & 
    \colhead{Decl.} & 
    \colhead{$J$ } &
    \colhead{$H$ } &
    \colhead{$K$ } &
    \colhead{$\Delta J$ } &
    \colhead{$\Delta K$ } &
    \colhead{$\Delta J-H$ } &
    \colhead{$\Delta H-K$ } &
    \colhead{$S$}   \\
   \colhead{(J2000)} & 
    \colhead{(J2000)} & 
    \colhead{mag} &
    \colhead{mag } &
    \colhead{mag } &
    \colhead{mag} &
    \colhead{mag } &
    \colhead{mag } &
    \colhead{mag} &
    \colhead{~}  
  }
\startdata
  20:57:33.38 & +52:02:51.9 & 15.65 & 14.81 & 14.34 & 0.18 & 0.22 & 0.13 & 0.19 & 2.12 \\ 
  20:57:33.82 & +52:16:57.1 & 14.04 & 13.51 & 13.30 & 0.10 & 0.09 & 0.07 & 0.09 & 1.04 \\ 
  20:57:47.82 & +52:35:24.6 & 13.35 & 12.60 & 12.23 & 0.07 & 0.08 & 0.10 & 0.07 & 1.16 \\ 
  20:57:51.99 & +52:52:33.7 & 13.91 & 13.08 & 12.75 & 0.05 & 0.08 & 0.05 & 0.08 & 1.04 \\ 
  20:58:08.27 & +52:22:24.4 & 16.39 & 15.48 & 15.07 & 0.13 & 0.11 & 0.14 & 0.10 & 1.30 \\ 
  20:58:18.83 & +52:13:07.4 & 15.68 & 14.65 & 14.20 & 0.10 & 0.14 & 0.09 & 0.07 & 1.60 \\ 
  20:58:18.92 & +52:10:10.1 & 16.26 & 15.44 & 15.13 & 0.12 & 0.14 & 0.12 & 0.12 & 1.18 \\ 
  20:58:22.06 & +52:26:47.8 & 14.83 & 14.18 & 13.90 & 0.07 & 0.12 & 0.07 & 0.09 & 1.07 \\ 
  20:58:27.66 & +52:18:31.9 & 15.55 & 14.47 & 13.94 & 0.11 & 0.05 & 0.09 & 0.07 & 1.57 \\ 
  20:58:35.47 & +52:42:41.0 & 14.05 & 13.06 & 12.65 & 0.11 & 0.06 & 0.09 & 0.07 & 1.11 \\ 
  20:58:35.74 & +52:43:53.1 & 15.20 & 14.59 & 14.26 & 0.12 & 0.09 & 0.09 & 0.07 & 1.26 \\ 
  20:58:37.31 & +52:28:08.2 & 16.91 & 15.52 & 14.82 & 0.23 & 0.14 & 0.20 & 0.12 & 1.64 \\ 
  20:58:41.61 & +52:42:45.8 & 15.62 & 14.64 & 14.24 & 0.14 & 0.07 & 0.14 & 0.10 & 1.02 \\ 
  20:58:45.02 & +52:42:59.0 & 15.39 & 14.63 & 14.37 & 0.17 & 0.10 & 0.10 & 0.10 & 1.28 \\ 
  20:58:46.47 & +52:29:41.6 & 13.20 & 11.83 & 11.07 & 0.20 & 0.12 & 0.10 & 0.10 & 1.34 \\ 
  20:58:46.95 & +52:08:23.5 & 14.70 & 14.09 & 13.87 & 0.09 & 0.08 & 0.10 & 0.13 & 1.28 \\ 
  20:58:47.02 & +52:41:40.6 & 16.69 & 16.13 & 15.62 & 0.97 & 0.57 & 1.01 & 0.44 & 1.35 \\ 
  20:58:48.88 & +52:22:31.6 & 16.89 & 14.86 & 13.77 & 0.22 & 0.07 & 0.19 & 0.10 & 1.38 \\ 
  20:58:52.78 & +52:19:39.1 & 15.98 & 15.24 & 14.95 & 0.10 & 0.11 & 0.10 & 0.10 & 1.20 \\ 
  20:58:53.16 & +52:42:46.7 & 14.88 & 14.07 & 13.72 & 0.16 & 0.09 & 0.11 & 0.09 & 1.05 \\ 
  20:59:03.46 & +52:02:33.9 & 15.16 & 14.27 & 13.92 & 0.08 & 0.15 & 0.06 & 0.10 & 1.18 \\ 
  20:59:04.29 & +52:33:02.6 & 16.94 & 14.21 & 12.78 & 0.21 & 0.05 & 0.19 & 0.06 & 1.05 \\ 
  20:59:11.42 & +52:46:41.9 & 14.82 & 13.87 & 13.45 & 0.11 & 0.08 & 0.07 & 0.07 & 2.20 \\ 
  20:59:12.17 & +52:09:42.0 & 14.78 & 14.21 & 14.02 & 0.08 & 0.07 & 0.05 & 0.05 & 1.07 \\ 
  20:59:13.54 & +52:03:59.1 & 16.45 & 15.20 & 14.65 & 0.16 & 0.10 & 0.14 & 0.10 & 1.27 \\ 
  20:59:17.76 & +52:46:12.3 & 15.75 & 14.81 & 14.36 & 0.11 & 0.09 & 0.11 & 0.07 & 1.80 \\ 
  20:59:22.91 & +52:35:13.8 & 14.55 & 13.63 & 13.25 & 0.08 & 0.08 & 0.08 & 0.12 & 1.01 \\ 
  20:59:32.26 & +52:32:47.4 & 13.93 & 13.31 & 12.97 & 0.07 & 0.08 & 0.05 & 0.05 & 1.21 \\ 
  20:59:33.88 & +52:10:16.8 & 13.15 & 12.67 & 12.60 & 0.07 & 0.07 & 0.04 & 0.05 & 1.11 \\ 
  20:59:41.00 & +52:23:49.2 & 16.98 & 15.26 & 14.23 & 0.23 & 0.06 & 0.21 & 0.10 & 1.26 \\ 
  20:59:49.14 & +52:04:42.7 & 16.53 & 15.39 & 14.94 & 0.13 & 0.09 & 0.11 & 0.09 & 1.13 \\ 
  20:59:57.42 & +52:38:57.2 & 15.82 & 14.20 & 13.36 & 0.09 & 0.04 & 0.09 & 0.11 & 1.20 \\ 
  20:59:59.71 & +52:10:43.1 & 16.17 & 15.34 & 15.10 & 0.12 & 0.12 & 0.13 & 0.14 & 1.14 \\ 
  21:00:01.78 & +52:25:45.3 & 16.56 & 14.98 & 14.10 & 0.17 & 0.07 & 0.17 & 0.09 & 1.11 \\ 
  21:00:13.31 & +52:02:24.6 & 16.30 & 15.24 & 14.69 & 0.14 & 0.18 & 0.19 & 0.19 & 1.43 \\ 
  21:00:15.00 & +52:30:06.0 & 14.07 & 13.43 & 13.30 & 0.12 & 0.10 & 0.08 & 0.07 & 1.18 \\ 
  21:00:19.16 & +52:27:16.0 & 17.15 & 15.13 & 14.02 & 0.20 & 0.07 & 0.20 & 0.08 & 1.05 \\ 
  21:00:20.48 & +52:31:08.4 & 14.19 & 12.70 & 11.89 & 0.11 & 0.13 & 0.08 & 0.08 & 1.14 \\ 
  21:00:22.53 & +52:42:53.7 & 15.94 & 15.14 & 14.79 & 0.27 & 0.17 & 0.14 & 0.15 & 1.43 \\ 
  21:00:22.81 & +52:07:33.7 & 15.86 & 14.96 & 14.68 & 0.11 & 0.12 & 0.12 & 0.13 & 1.48 \\ 
  21:00:23.12 & +52:04:32.7 & 16.10 & 14.59 & 13.77 & 0.12 & 0.10 & 0.12 & 0.06 & 1.03 \\ 
  21:00:23.32 & +52:42:58.4 & 15.79 & 14.62 & 14.12 & 0.21 & 0.13 & 0.12 & 0.13 & 1.47 \\ 
  21:00:23.49 & +52:54:18.2 & 15.83 & 14.91 & 14.50 & 0.11 & 0.12 & 0.12 & 0.10 & 1.10 \\ 
  21:00:23.97 & +52:42:20.5 & 15.09 & 14.00 & 13.55 & 0.18 & 0.15 & 0.11 & 0.12 & 1.05 \\ 
  21:00:37.50 & +52:07:04.1& 16.88 & 15.65 & 15.03 & 0.60 & 0.12 & 0.60 & 0.13 & 2.40 \\ 
  21:01:03.61 & +52:20:42.8 & 15.58 & 14.56 & 14.13 & 0.11 & 0.09 & 0.06 & 0.06 & 1.45 \\  
  21:01:06.83 & +52:04:32.1 & 17.11 & 16.27 & 15.79 & 0.76 & 0.23 & 0.71 & 0.22 & 2.15 \\ 
  21:01:13.48 & +52:09:01.3 & 14.72 & 13.92 & 13.54 & 0.07 & 0.10 & 0.04 & 0.05 & 1.22 \\ 
  21:01:36.23 & +52:21:44.2 & 13.61 & 12.86 & 12.46 & 0.09 & 0.15 & 0.07 & 0.11 & 2.62 \\ 
  21:01:41.63 & +52:02:17.7 & 15.17 & 13.60 & 12.90 & 0.19 & 0.14 & 0.18 & 0.11 & 1.10 \\ 
  21:01:55.68 & +52:26:59.6 & 17.21 & 14.07 & 12.42 & 0.26 & 0.07 & 0.23 & 0.08 & 1.26 \\ 
  21:01:57.61 & +52:04:30.3 & 16.87 & 15.85 & 15.44 & 0.24 & 0.17 & 0.27 & 0.23 & 1.25 \\ 
  21:02:06.63 & +52:07:31.0 & 16.32 & 15.31 & 14.70 & 0.11 & 0.12 & 0.16 & 0.16 & 1.02 \\ 
  21:02:17.82 & +52:07:31.5 & 16.87 & 15.97 & 15.61 & 0.18 & 0.42 & 0.38 & 0.29 & 1.40 \\ 
  21:02:19.35 & +52:04:46.1 & 13.57 & 12.85 & 12.55 & 0.07 & 0.08 & 0.05 & 0.09 & 1.40 \\ 
  21:02:34.21 & +52:09:33.8 & 14.66 & 13.76 & 13.43 & 0.07 & 0.07 & 0.06 & 0.05 & 1.30 \\ 
  21:03:14.00 & +52:03:37.9 & 16.85 & 15.40 & 14.77 & 0.30 & 0.19 & 0.24 & 0.13 & 1.09 \\ 
  21:03:14.92 & +52:03:58.5 & 14.66 & 13.90 & 13.49 & 0.09 & 0.16 & 0.12 & 0.13 & 1.12 \\ 
  21:03:17.92 & +52:16:38.0 & 12.91 & 12.41 & 12.34 & 0.13 & 0.14 & 0.07 & 0.11 & 1.04 \\ 
  21:03:18.27 & +52:17:46.6 & 15.48 & 14.23 & 13.64 & 0.09 & 0.14 & 0.08 & 0.09 & 1.07 \\ 
  21:03:18.42 & +52:03:36.9 & 15.55 & 14.12 & 13.46 & 0.11 & 0.18 & 0.17 & 0.16 & 1.00 \\ 
  21:03:18.56 & +52:15:06.5 & 14.09 & 13.38 & 13.11 & 0.15 & 0.14 & 0.11 & 0.09 & 1.09 \\

  \enddata

  \tablecomments{Typical photometric errors are $\sim2\%$}

\end{deluxetable}



\begin{deluxetable}{lllllllccccc}

\tabletypesize{\scriptsize}
  
  \tablecaption{Other possibly periodic stars in the monitored field.}
  \label{Table:weak}
  \tablewidth{0pt}

 \tablehead{
     \colhead{R.A. } & 
    \colhead{Decl.} & 
    \colhead{$J$ } &
    \colhead{$H$ } &
    \colhead{$K$ } &
    \colhead{$\Delta J$ } &
    \colhead{$\Delta K$ } &
    \colhead{$\Delta J-H$ } &
    \colhead{$\Delta H-K$ } &
    \colhead{$S$}  &
    \colhead{Period}&
\colhead{Gr.} \\

   \colhead{(J2000)} & 
    \colhead{(J2000)} & 
    \colhead{mag} &
    \colhead{mag } &
    \colhead{mag } &
    \colhead{mag} &
    \colhead{mag } &
    \colhead{mag } &
    \colhead{mag} &
    \colhead{~}  &
\colhead{(days)} &
 \colhead{$1-5$}  }
\startdata
{\bf Long} & {\bf period}  &  &  &  &  &  &  &  & &   &  \\
\hline
  20:59:41.783 & +52:39:10.2 & 16.52 & 15.22 & 14.52 & 0.40 & 0.28 & 0.20 & 0.16 & 5.30 & 61 or 122* &3 \\ 
  21:00:00.453 & +52:24:50.5 & 13.92 & 12.42 & 11.72 & 0.19 & 0.15 & 0.07 & 0.07 & 3.16 & 59.20 & 3 \\ 
  21:02:22.239 & +52:44:57.9 & 15.53 & 14.23 & 13.60 & 0.17 & 0.10 & 0.16 & 0.07 & 2.04 & 72.60 & 2 \\  
{\bf Moderate} & {\bf period}  &  & &  &  &  &  &  &  &  \\
\hline  
  20:57:47.04 & +52:43:57.6 & 14.74 & 13.88 & 13.50 & 0.10 & 0.17 &    0.10 & 0.09 & 1.12 & 2.78 & 4    \\  
  20:57:51.19 & +52:31:54.6 & 14.92 & 13.75 & 13.21 & 0.13 & 0.10 & 0.09 & 0.05 & 2.30 & 17.85 & 5	 \\  
  20:58:11.02 & +52:26:53.4 & 15.58 & 14.32 & 13.68 & 0.09 & 0.08 & 0.06 & 0.07 & 1.02 & 12.43 & 3	 \\  
  20:58:16.96 & +52:47:34.7 & 12.30 & 11.95 & 11.75 & 0.08 & 0.09 & 0.07 & 0.07 & 1.83 & 12.80 & 2	 \\  
  20:58:19.12 & +52:21:40.6 & 14.83 & 13.72 & 13.20 & 0.12 & 0.09 & 0.09 & 0.07 & 2.31 & multi* & 2	 \\  
  20:58:23.86 & +52:20:10.1 & 13.33 & 12.80 & 12.47 & 0.07 & 0.07 & 0.06 & 0.08 & 1.50 & 9.84 & 4	 \\   
  20:58:27.49 & +52:36:42.6 & 16.58 & 15.28 & 14.61 & 0.19 & 0.10 & 0.21 & 0.09 & 1.03 & 25.80 & 2	 \\   
  20:58:30.78 & +52:06:39.0 & 14.10 & 13.42 & 13.16 & 0.12 & 0.09 & 0.08 & 0.10 & 1.68 & 4.52 & 4	 \\  
  20:58:42.08 & +52:20:03.5 & 13.95 & 13.03 & 12.65 & 0.12 & 0.08 & 0.09 & 0.08 & 1.36 & 4.70 & 4	 \\   
  20:59:17.76 & +52:46:12.3 & 15.72 & 14.79 & 14.33 & 0.11 & 0.09 & 0.11 & 0.07 & 1.80 & 3.07 & 3	 \\  
  20:59:28.56 & +52:10:10.4 & 15.88 & 15.04 & 14.70 & 0.18 & 0.15 & 0.09 & 0.09 & 3.24 & 3.56 & 3	 \\   
  21:00:14.89 & +52:14:58.0 & 12.80 & 12.08 & 11.90 & 0.09 & 0.07 & 0.10 & 0.10 & 1.62 & 1.25,  4.96*  & 1\\  
  21:00:16.18 & +52:47:59.6 & 16.28 & 15.39 & 15.00 & 0.16 & 0.11 & 0.13 & 0.18 & 1.28 & 7.70 & 1	       \\  
  21:00:42.28 & +52:20:10.5 & 12.37 & 11.56 & 11.22 & 0.11 & 0.09 & 0.06 & 0.08 & 2.06 & 6.19 & 5	       \\  
  21:00:42.84 & +52:46:16.3 & 15.43 & 14.45 & 14.01 & 0.13 & 0.13 & 0.12 & 0.09 & 2.68 & 9.46 & 4	       \\ 
  21:01:21.18 & +52:43:43.7 & 16.01 & 15.13 & 14.73 & 0.24 & 0.10 & 0.25 & 0.12 & 1.52 & 4.7 , 3.1* & 3     \\  
  21:01:34.27 & +52:47:41.2 & 14.99 & 14.14 & 13.83 & 0.12 & 0.11 & 0.11 & 0.06 & 2.37 & 2.29$^\dag$  & 2   \\   
  21:02:02.70 & +52:39:29.0 & 15.74 & 14.55 & 14.00 & 0.12 & 0.09 & 0.07 & 0.08 & 1.06 & 12.50 & 1	       \\   
  21:02:16.70 & +52:02:19.9 & 15.23 & 14.30 & 13.96 & 0.30 & 0.10 & 0.31 & 0.14 & 1.72 & 4.59 & 3	       \\  
  21:02:35.59 & +52:43:53.6 & 14.38 & 13.46 & 13.07 & 0.17 & 0.13 & 0.09 & 0.08 & 3.64 & 7.29 & 5	       \\  
  21:03:02.06 & +52:53:49.4 & 15.98 & 15.12 & 14.84 & 0.22 & 0.15 & 0.24 & 0.12 & 1.26 & 8.11 & 3	       \\    
 \enddata
  \tablecomments{Typical photometric errors are $\sim2\%$.}
  \tablenotetext{*}{Noted stars did not have a unique period identified.}
   \tablenotetext{\dag}{Period only noted in Season 2.}
\end{deluxetable}

\subsection{Candidate Eclipsing Systems}	

	Eclipsing binaries are among the easiest periodic objects to detect due to their highly stable periodicity and their conspicuous sudden drops by 0.2 mag or more. 
Eclipsing binaries  can provide fundamental mass and radius measurements for the component stars \citep[see the extensive review by][]{And91}. These mass and radius measurements allow for accurate tests of stellar evolution models \citep[\eg ,][] {Pol97, Sch97, Gui00, Tor02}.  In their Kepler survey, \citet{Prs11} identify 1879 such objects in the 105$^{\rm o}$$^2$ field. 
There are over 100 variability types listed in the Variable Star Catalog (GCVS)  \citep{Sam09}.  It is not our purpose to precisely identify each variable with its exact type;
however, their short periods,  large amplitudes and stability generally rule out most varieties of pulsating stars.
Because our observations span a year and a half, we can highly constrain the shapes of these stars' periodic light curves despite not observing with a cadence of more than once per night.


Among the 149 diskless variable stars, we identified about 60 candidate eclipsing binaries.  The determining factors for the assignment of eclipsing as the reason for the variation were a) strong period detections which were consistent between the seasons with no phase shift and b) generally short periods. 
We visually inspected each folded light curve and found  51 of these with very clear periods, consistent, season to season and filter to filter.  Since their periods tended to be short, usually F$\chi^2$ fitting was more appropriate.  
  The GCVS divides eclipsing binary systems into four  observational classes based on the shape of their light curves.  
  The general classes are: 1) `E' for eclipsing systems in which the stars are well separated with very sharp eclipses.
2)   `EA' for Algol-like system in which we can notice some rounding of the edges near the time of the eclipses.
3)   `EB' for $\beta$ Lyrae-like system in which the eclipse edges are so rounded as to make it difficult to tell when the eclipses begin and end, and they have one minimum much deeper than the other. 
4)   `EW' for W UMa-like near contact systems with nearly sinusoidal signals and period $\lesssim$ 1 day. 
We note these classes in the last column of Table~3. 
 There are a few stars in Table~3 identified as pulsators; these are usually identified with `P' except for one suspected RR~Lyrae (`RR').
We do not attempt to classify the lightcurves by the physical characteristics of the stars as we lack sufficient data.
 
 Periods for the remaining stars were less certain due to either lower signal strength or aliasing which made it difficult to select the correct period out of a few candidates. Usually this was resolved by using the $\sim$ 500 day photometric base-line.  In some cases, the signal levels were close to the noise, so single-peaked and double-peaked signals were difficult to distinguish.  While some of the reported periods are below our formal sensitivity for periods given the sampling of our data, the period-folded lightcurves are convincing enough that we are confident enough to include them.  
We do not make classifications for Table~5 as the quality of the folded lightcurves was not high enough.

\begin{figure}
   \includegraphics[scale=1.0]{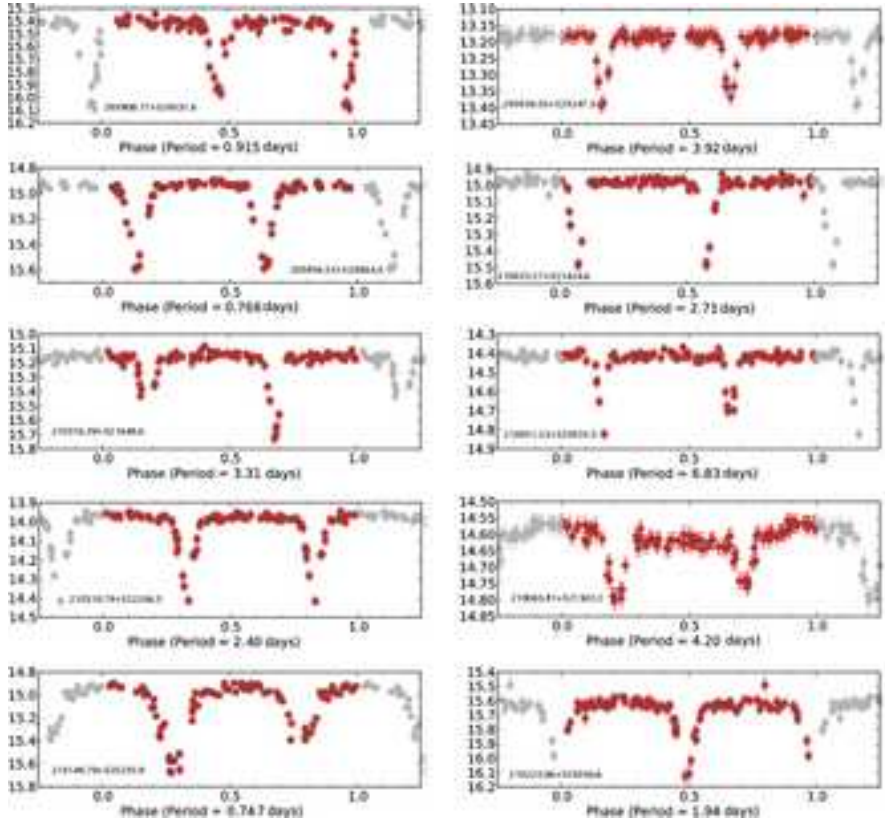}

     \caption{
Some examples of detached systems.  $K$-band data are shown. Phase 0 is arbitrary based on the time of the first observation.
Names are indicated in each panel. In the bottom two rows, ellipsoidal effects are clearly seen. 
}
  \label{Fig:SE}
\end{figure}

 
 Just under 40\% of the eclipsing systems showed clean, sharp eclipses, in which the star deviated from its nominal flux level for a fixed time period in which it dropped sharply and then returned to the original flux.   These are detached systems  where the separation of both components is large compared to their radii. The stars interact gravitationally, but the distortion of their surfaces due to tidal deformation and rotation is minimal.  
 These constitute the first group of stars in Table~3, 
 identified as ``simple eclipse''. These systems have periods ranging from around 0.4 days to a little over 13 days.  
For the best examples, these stars ranged from an average J magnitude of 13.3 to 16.8 mag with a median average $J\sim 16.2$.
Eclipse depth ranges from 0.15 to 0.92 mag with a median of 0.45 mag.  The color changes are small, but measurable, with the median shift in $J-K$ color of about 0.18 and and a maximum shift of about $\Delta J-K \sim 0.5$. We show 10 of these in Figure~\ref{Fig:SE}. The nominal level when the system is out of eclipse indicates limited tidal distortion effects -- with  a few exceptions, 210149.79+525235.9 being the most conspicuous. 

\subsection{Ellipsoidal/Contact systems} 

Slightly over 60\%  of the lightcurves in this sample had  continuous flux variations.  
There have been several recent surveys focused on finding variable stars.  A long baseline study covering up to 8 years is a galactic field monitoring program by Paczy\'nski et al. (2006). It is more common that these studies last an observing season (\eg Weldrake \& Bayliss 2008 and Miller et al. 2010). \citet{Pie09} and \citet{Pie11} focused on deep monitoring program of a few days which included NIR and optical photometry.  The results of these programs are that between 0.06 and 0.30 \% of the field stars show systems with continuous flux variations, indicative of contact binaries or elliptical stars.  An exception is the result from the $Kepler$ mission which indicates the rate of such stars may be as high as 1.2\% when the precision in the photometric is high enough \citep[\eg mmag,][]{Prs11}.

\begin{figure}
                 \includegraphics[scale=1.0]{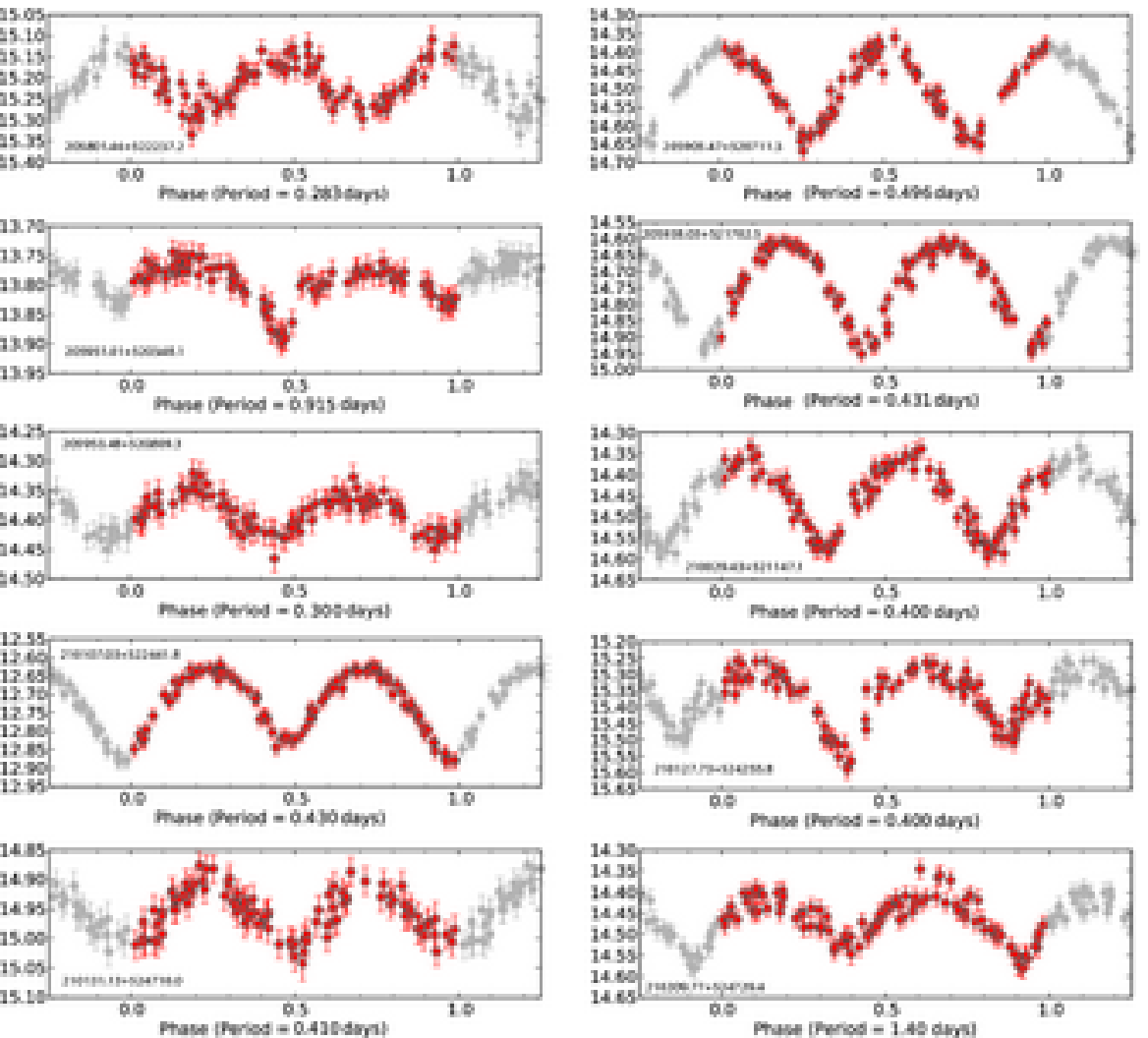}

  \caption{
Some example contact systems. Names are indicated in each panel.
}
  \label{Fig:OC}
\end{figure}

While it is possible the continuous signal variation is due to the stars being ellipsoidal in shape,  it is more likely that two stars are in or near contact with each other and there is a continuous variation of flux.   In contact binary systems all cycles contain two maxima and two minima and this was enforced by the period finding algorithm.   We show 10 of these in Figure~\ref{Fig:OC}.
We would expect that the periods of these contact binaries are shorter than other binaries.  Indeed the periods range from about 0.2 days to 3 days for these kinds of systems. The median period found was around 0.4 days.   Periodic phenomena are also expected in young spotted stars, but the short periods and and long term stability argue against this interpretation.  Some of the qualitatively weaker sources, such as 205922.81+522338.4, 
  205746.71+520302.3 and  
  210058.93+523810.7, 
  were explicitly examined at high order peaks in their periodograms, but no convincing signal was found.  The periods shown in Figure~\ref{Fig:OC} here are the strongest periods for these stars, but the two algorithms occasionally returned alias frequencies.   These stars ranged from an average J magnitude of 12.5 to 17.3 with a median of  $J\sim 16.1$.
Eclipse depth ranges from $\Delta J =$ 0.1 to 0.9 with a median of 0.35. 
 The color changes are small but measurable  with the median shift in $J-H$ color of about 0.2 and and a maximum shift of about $\Delta J-H \sim 0.4$. 
A few systems had clear color patterns.    205945.69+520716.3 
tends to become redder as it becomes fainter and bluer as it become brighter;    210029.43+521147.1 
showed a similar pattern.   205905.47+520711.3 and  205746.71+520302.3 
appear reddest when they are faintest, but the effect is weak, perhaps 0.02 in $H-K$. A detailed statistical analysis would be needed to prove the color patterns noted in these stars were significant.

 Most of the contact systems are probably W~UMa type systems,
consisting of a pair of stars in a tight, circular orbit, with each star in contact with and eclipsing the other. 
W~UMa type contact binary stars are surprisingly common in the solar neighborhood. The apparent relative frequency of W~UMa type stars was estimated at about 1/130 of the frequency of the common
solar-neighborhood FGK-type disc population dwarfs with an
age of $5-11 \times 10^9$  years \citep[OGLE][]{Ruc98}.
This is similar to the detection rate found here, which was about  32/9200 ($\approx$1/240 or 0.04\%)  
Because the formation of contact systems by the gradual loss of angular momentum due to
the combined effects of tidal coupling and braking by the stellar
wind W~UMa type systems are not observed in clusters younger than  about 1 billion years old.

\subsection{Longer period stars}
About 20 stars showed sinusoidal variations on timescales of about 2 to 25 days. These are indicated as ``moderate period/ weak signal'' variables in Table~5.
Many of these show significantly more deviation from the computed signal than the typical short period system.  
This is not an effect of low peak to trough signal (and hence higher noise) -- the best examples of stars with period from 2-25 days ranged from and average J magnitude of 12.4 to 15.5 with a median average $J\sim 14.1$ - brighter than either of the eclipsing subgroups. 

\begin{figure}
      \includegraphics[scale=1.0]{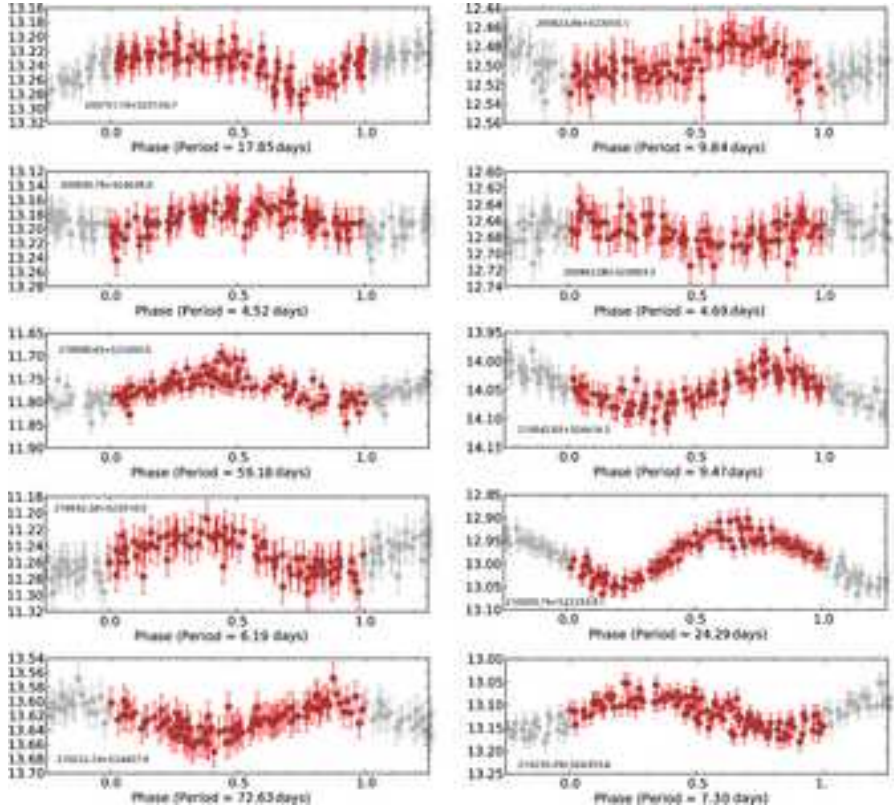}

  \caption{
  Three--season $K$ band period-folded lightcurves of stars with quasi-sinusoidal periods longer than 2 days which were stable across the three seasons  
   including 2 long period variables. Names are indicated in each panel. 
  \label{Fig:slow rotator}}
\end{figure}

The confidence of most of the periods in the lower portion of Table~5 is not very high as noted by the relatively low grades. 
We show 10 examples of these in Figure~\ref{Fig:slow rotator}.  Most of these objects have light curves dominated by a single peak and single trough. Unlike the previous group of objects, doubling the periods did not improve the signal to noise. Several stars in the group showed periodicity in only some of the observing seasons. Other stars showed several signals at similar strength and we chose to list the strongest two.  One star,  205819.12+522140.6, has five equally probable signals.  This might be a sign that some of these are pulsating stars, such stars would show relatively colorless changes as observed here. 

Another possibility is that some of these stars may be diskless, low-mass YSOs, alternatively called Class~III objects or weak T~Tauri stars.  As this is a region of star formation with about 40 previously identified disk and envelope bearing stars, it would be surprising if the number of diskless objects was much less than 20; it could possibly exceed 40. Diskless YSOs are often variable due to surface spots induced by dynamo behavior in their evolving interiors \citep[\eg][]{Fei99}.

Cool spots, like those on the Sun, were first identified as a contributor to the variability of PMS stars in the 1980s \citep{Vrb85}.
Static stellar spots induce variability because of the rotation of the star. Starspots have been used  regularly as a method of measuring stellar periods \citep[e.g.][]{Att92}.   Typical rotation periods of  weak T~Tauri stars are between 1 and 10 days \citep{Gra08}.  In the I band, the luminosity change is typically $<$ 15\% \citep{Coh04}.  
The implied color change due to a lower effective temperature is $<$ 5\%.  
The median magnitude change of the moderate period stars in our data is  $\sim$ 10\%, quite consistent with the spot interpretation. 
 Typical color changes are a bit higher than expected,  about 8\% at $J-H$ and $H-K$,  however color measurements are more sensitive to measurement precision so we do not consider this result inconsistent with the spot interpretation -- at least for some of the sources.

\subsection{Other periodic stars}
\begin{figure}
 \includegraphics[scale=1.0]{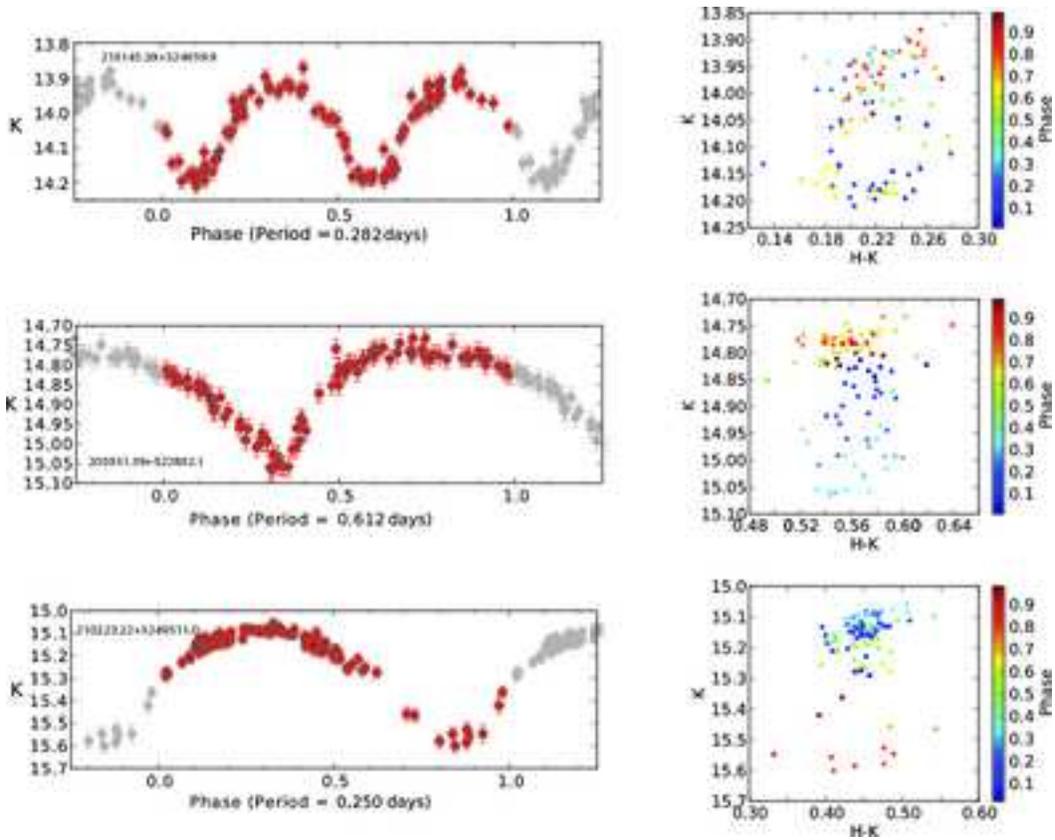}
 
   \caption{Three stars with very short, but stable periods. Names are given in each panel.
 Top-to-bottom are: an eclipsing system, an  RRab star and a probable $\delta$ Scuti star.
   For each the left panel is the full $K$-band data set  folded on the period. 
    The right panels show the $K$ vs. $H-K$ color magnitude diagrams.}
     \label{pulse}
\end{figure}

 	A small number of stars showed very strong periodic light curves which neither corresponded to eclipse events or symmetric signals.  
	These are indicated as ``other periodic'' stars in Table 3. Two examples are shown in Figure~\ref{pulse}.   Star  210145.39+524059.9 
	has a period of just 6.8 hours with a one flat minima which lasts about an hour   $\Delta J \sim \Delta K \approx 0.33$.  There is a clear pattern in the color-magnitude diagram indicating the star changes color, becoming bluer the longer it is in the faint state, reddening somewhat as it brightens and then becoming fainter with little noticeable color change.  This appears to be an eclipsing binary.
 Star 205931.39+522802.1 
 has a period of about 14.7 hours with a smooth rapid rise coming about 1/3 of the cycle and the decline lasting the other 2/3
 $\Delta J =0.48$  and $\Delta J-K \sim 0.15$.   This appears to be an RR Lyrae sub-type known as an RRab star which has a large amplitude, hours long, asymmetric light curve. 
 Star  210223.22+524951.0 
 shows a combination of these effects: a short ($\sim 6$ hour) period, a sustained minimum, a rapid rise and a slow fall.   For this star $\Delta J =0.64$  and $\Delta J-K \sim 0.10$.   This may be a $\delta$ Scuti star.

There are a few stars which appear to vary on longer times scales.  Because there are only three, and the periods are at least of order the duration of the observing season, it is hard to tell if they share any characteristics.
There are also a few objects with very regular, non-sinusoidal variations.  Some of these may instead be tidally distorted stars or perhaps pulsating stars. 

Since our period detection algorithms were designed to find stable periods we may have missed  $\delta$ Scuti stars - these are among the most common pulsating stars \citep{Pie09}.
One of the key distinguishing characteristic of common pulsating stars is irregular change signal strength on a regular period. 
 The typical $\delta$ Scuti star has an amplitude of 0.003 - 0.9 magnitudes. The variations in luminosity are due to both radial and non-radial pulsations of the star's surface. They sometimes have beat patterns which make the light curves difficult to fold as the amplitude of adjoining cycles is often modulated.  Further the periods are typically $<$ 6 hours which is much below our sampling.   $\delta$ Scutis are early A to late F range and hence are generally warm with respect to the near-IR studied here, with color changes more obvious in optical wavelengths. The J-K color shouldn't change much at all.
Given the 0.2\% fraction observed in Carina \citep{Pie09} we expect about 18 of these stars in our dataset; however, given their typical periods of 0.25 days and typical amplitudes, including amplitude modulation, many of these may have gone undetected or noted as stochastic systems.


\section{Conclusions} 
\label{sec:conclusion}
In this contribution, we present the results of an approximately 100 night, NIR photometric monitoring campaign. While the program itself was focused on the pre-main sequence population of the very young star forming region Cyg~OB7,  we report here on the field star population. 
The $\sim$ 9,200 non-disked stars surveyed were in a band from $11<J<17.3$, $11<H<16.5$, and $11<K<16$. 
About 1.6\% of the non-disked stars variable.  This is significantly less than the rate for disked YSO's which have variability rates $>$ 90\%
\citep{Car01, Car02, Mor09, Mor11, Ric12, Sch12}.  Of the variables, about 1/3 are weakly variable with stochastic changes.  

We used two period finding algorithms on the variables,  Fast $\chi^2$ and the Lomb-Scargle periodogram.    About one-third of the variables are clearly eclipsing systems with highly stable periods.    The greater half of the eclipsing systems are contact or near contact binaries with continous flux changes.  The detection rate of contact/ellipsoidal binary systems is 0.4\%, about 25\% higher than other ground-based surveys, but about one-third the rate found by {\it Kepler}.

Among the remaining variables, some of these have very stable, very short periods and appear to be pulsators. There are intermediate length periodic variables with less stable variability, with  nearly colorless luminosity changes of about 0.1 magnitudes. 
 These stars have periods between 2 and 25 days and may be spotted. Follow-up deep X-ray or optical H$\alpha$ observations could be used to ascertain the evolutionary status of these stars and the likelihood that these changes do indeed arise from spots.

\section{Acknowledgements}
 S.J.W. is supported by NASA contract NAS8-03060 (Chandra). T.S.R.  was supported by Grant \#1348190 from the Spitzer Science Center and also the NSF REU site grant \#0757887.
 The United Kingdom Infrared Telescope is operated by the Joint Astronomy Centre on behalf of the Science and Technology Facilities Council of the U.K. We thank A. Nord, L. Rizzi, and T. Carroll for assistance in obtaining these observations. We also thank the University of Hawaii Time Allocation Committee for allocating the nights during which these observations were made.
 The authors wish to recognize and acknowledge the very significant cultural role and reverence that the summit of Mauna Kea has always had within the indigenous Hawaiian community. We are most fortunate to have the opportunity to conduct observations from this sacred mountain. Special thanks to Nancy Evans, Margarita Karovska and Matthew Templeton for helping us assess the lightcurves.

\end{document}